\documentclass[a5paper,graybox]{svmult}

\usepackage{type1cm}        
\usepackage{makeidx}         
\usepackage{graphicx} 
\usepackage{multicol} 
\usepackage[bottom]{footmisc}

\usepackage{geometry}
\usepackage {newtxtext} %
\usepackage {newtxmath} 
\usepackage{graphicx}
\graphicspath{{Figures/}}

\usepackage{xcolor}         
\definecolor{mylinkcolor}{HTML}{20639b}

\definecolor{mylinkcolor2}{HTML}{f8f8f7}
\definecolor{redcolor}{HTML}{eae2b7}

\usepackage{hyperref}
\hypersetup{
    colorlinks=true,
    linkcolor=mylinkcolor,  
    citecolor=mylinkcolor,  
    urlcolor=mylinkcolor,   
}
\makeatletter
\renewcommand\@cite[2]{%
    [{\hypersetup{citecolor=mylinkcolor}[{#1\if@tempswa , #2\fi}]}]
}
\makeatother

\usepackage{tikz,xcolor,hyperref}
\definecolor{lime}{HTML}{A6CE39}
\DeclareRobustCommand{\orcidicon}{%
    \begin{tikzpicture}
    \draw[lime, fill=lime] (0,0) 
    circle [radius=0.16] 
    node[white] {{\fontfamily{qag}\selectfont \tiny ID}};
    \draw[white, fill=white] (-0.0625,0.095) 
    circle [radius=0.007];
    \end{tikzpicture}
    \hspace{-2mm}
}
\newcommand{\orcid}[1]{\href{https://orcid.org/#1}{\orcidicon}}

\usepackage{everypage}  
\AddEverypageHook{%
  \begin{tikzpicture}[remember picture, overlay]
    \node[anchor=south, xshift=0mm, yshift=1mm] at (current page.south) {
      \begin{tikzpicture}
        \node[fill=redcolor, text=black, rounded corners=0mm, inner sep=3mm]{
          \scriptsize \textit{Contributed chapter to "Nonlinear Dynamics for Biological Systems", M. Stich, J. Carballido-Landeira (Eds), Springer, Switzerland, 2024}
        };
      \end{tikzpicture}
    };
  \end{tikzpicture}%
}
\usepackage[numbers]{natbib}

\begin{document}

\title*{Networks: The Visual Language of Complexity}

\author{Blai Vidiella, Salva Duran-Nebreda and Sergi Valverde}

\institute{
\textbf{Blai Vidiella \orcid{0000-0002-4819-7047} : } $^1$Institute of Evolutionary Biology, CSIC-UPF, Pg. Barceloneta 37, Barcelona 08003, Spain. $^2$Theoretical and Experimental Ecology Station, CNRS, Moulis, France.\\
\email{blai.vidiella-rocamora@sete.cnrs.fr}\\
\textbf{Salva Duran-Nebreda \orcid{0000-0002-2539-3539} : } $^1$Institute of Evolutionary Biology, CSIC-UPF, Pg. Barceloneta 37, Barcelona 08003, Spain. \email{salva.duran@ibe.upf-csic.es}\\
\textbf{Sergi Valverde \orcid{0000-0002-2150-9610} : } $^1$Institute of Evolutionary Biology, CSIC-UPF, Pg. Barceloneta 37, Barcelona 08003, Spain. 
$^3$European Centre for Living Technology (ECLT), Ca’ Bottacin, Dorsoduro 3911, 30123 - Venice, Italy. \email{s.valverde@csic.es}
}

%
%
\maketitle

\abstract{Understanding the origins of complexity is a fundamental challenge with implications for biological and technological systems. Network theory emerges as a powerful tool to model complex systems. Networks are an intuitive framework to represent inter-dependencies among many system components, facilitating the study of both local and global properties. However, it is unclear whether we can define a universal theoretical framework for evolving networks. While basic growth mechanisms, like preferential attachment, recapitulate common properties such as the power-law degree distribution, they fall short in capturing other system-specific properties. Tinkering, on the other hand, has shown to be very successful in generating modular or nested structures `for-free', highlighting the role of internal, non-adaptive mechanisms in the evolution of complexity. Different network extensions, like hypergraphs, have been recently developed to integrate exogenous factors in evolutionary models, as pairwise interactions are insufficient to capture environmentally-mediated species associations. As we confront global societal and climatic challenges, the study of network and hypergraphs provides valuable insights, emphasizing the importance of scientific exploration in understanding and managing complexity.}


\section{Introduction}\label{sec:intro}

\begin{figure}[hb]
    \centering
    \includegraphics[width=0.98\textwidth]{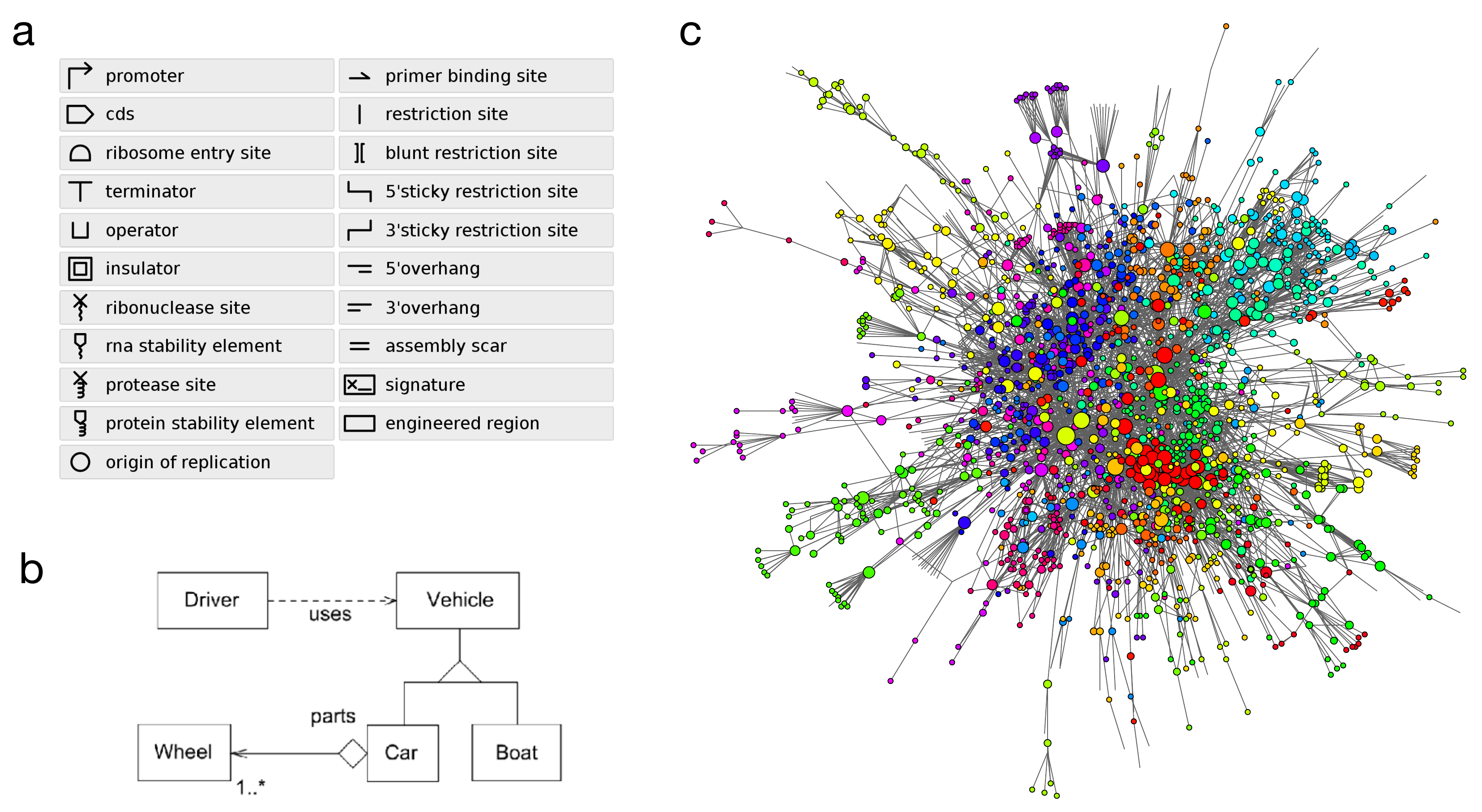}
    \caption{{\bf Visual notations in biology and software.} (a) The Synthetic Biology Open Language (SBOL) defines standardised visual notations for representing genetic constructions and biological designs such as DNA sequences, regulatory elements, and functional parts, as well as their connections and interactions within a biological system. Visual symbols represent promoters, coding sequences, terminators, and other genetic components, making it easier to communicate biological ideas across communities. (b) In software engineering, class diagrams are a form of structural diagram implemented in the Unified Modeling Language (UML). They serve the purpose of conceptualizing and representing the distinct attributes of software components, as well as their interrelationships, within a given system or software application. The illustration depicts a section of the class diagram used to create a racing game. (c) The network representation of the class diagram in the video game Prorally 2002 (Ubisoft, 2002).  Colour denotes the subsystem the software component belongs to (i.e., physics, rendering, audio, etc.), and node size represents the in-degree.}
    \label{fig:sbol}
\end{figure}

Understanding complexity is a main scientific challenge~\citep{bonner1988evolution}. This quest has been a complex process in and of itself, motivated by both theoretical and practical reasons. Consider two examples of using networks to comprehend complexity, one from biology and one from technology. Bertrand Russell proposed in the 1930s that notations play an important role in distilling abstract ideas into clearly intelligible forms, hence aiding communication~\citep{woodger1937axiomatic}. Almost 90 years later, the Synthetic Biology Open Language (SBOL)~\citep{madsen2019synthetic} introduced a formal notation in modern biology, aspiring to be an industrial standard for those engineering or simply conveying biological designs (particularly genetic constructs and pathways) (see Figure \ref{fig:sbol}a).\\

Prior to SBOL, computer scientists were seeking for a language that would allow them to control the complexity of software. Alan Kay, a former math and biology student and Xerox PARC computer scientist, recognized that the level of complexity in software was only comparable to that of biology. Kay played a significant role in the development of Smalltalk~\citep{kay1996early}, a programming language that conceptualized software as a network of code units, drawing inspiration from biological analogies and the interactions between cells: ``I thought of objects being like biological cells and/or individual computers on a network, only able to communicate with messages.'' \\

Smalltalk is a successful programming language, but neither it nor its contemporary descendants have been the ultimate solution to software development \citep{valverde2021long}. One of us participated in a video game project in 2002, where we used software engineering techniques, including graphical notation (see Figure \ref{fig:sbol}b). The large-scale structure of our video game proved to be extremely complex and difficult to understand, despite the talent of the software development team and ongoing attempts to optimize, improve, and modularize the software architecture \citep{valverde2002scale, valverde2007hierarchy} (see Figure \ref{fig:sbol}c). Unwanted complexity is certainly not surprising to many software developers, but neither is there a convincing explanation for it (or a strategy to prevent it).
For these reasons, we decided to launch an ongoing research project to investigate the origins of software complexity from a scientific (not engineering) perspective~\citep{valverde2005logarithmic}. This example also suggests that SBOL's comparable initiatives may have a limited impact in managing biological complexity (especially when dealing with large non-orthogonal biological designs~\citep{carbonell2014bottom,carbonell2016dealing}). But why are notation schemes insufficient? What is missing?\\

An important requirement is the ability to capture the structure-dynamics relationship as well as the evolutionary component inherent in complex systems. The crucial role of dynamics in complex systems is exemplified by social insect colonies, whose properties cannot be reduced to the agents building them~\citep{wilson2009superorganism}. Nest architecture emerges from the coordinated actions of many insects, which are, in turn, channeled through the emerging 3-D network of galleries~\citep{perna2008structure, valverde2009percolation}. Feed-backs are a universal feature of the evolution of complexity, which can be a driver of adaptations and innovations emerging at multiple scales~\citep{duran2017bridging}. Examples include not only social insects~\citep{wilson2009superorganism, sole2016synthetic}, but multicellularity~\citep{ratcliff2012experimental, brunet2017origin, duran2016emergence, sole2018road}, human language~\citep{corominas2009ontogeny} and many others~\citep{valverde2016major}. Evolutionary transitions frequently occur at critical points~\citep{szathmary2015toward,sole2011phase} that often involve (or can be understood by) some kind of structural property. Network theory is a tool for not only graphically displaying the structure of a complex system, but also for understanding the logic and evolutionary principles that underpin these systems. \\

Networks provide a means to investigate the relationship between system structure and dynamics, offering an intuitive framework to represent intricate inter-dependencies among system components. Their versatility lies in their ability to capture both local and global properties, enabling the analysis of small-scale interactions and large-scale emergent phenomena. Whether it's understanding the spread of diseases, analyzing information flow in social media, or optimizing transportation networks, graphs offer a robust mathematical language for studying complex systems and their dynamics. The effectiveness of networks as a mathematical representation of complexity has led researchers to embrace them in various scientific disciplines. This multidisciplinary approach facilitates the integration of theoretical and empirical approaches, allowing us to better comprehend and predict abrupt human-induced changes in ecosystems. Many examples demonstrate how networks can deepen our understanding of the impact of human actions on the biosphere.  By leveraging the power of networks, we can develop strategies to mitigate the effects of human actions, a crucial challenge in the era of climate change. 

\section{Measures}

\begin{figure}[hb]
    \centering
    \includegraphics[width=0.9\textwidth]{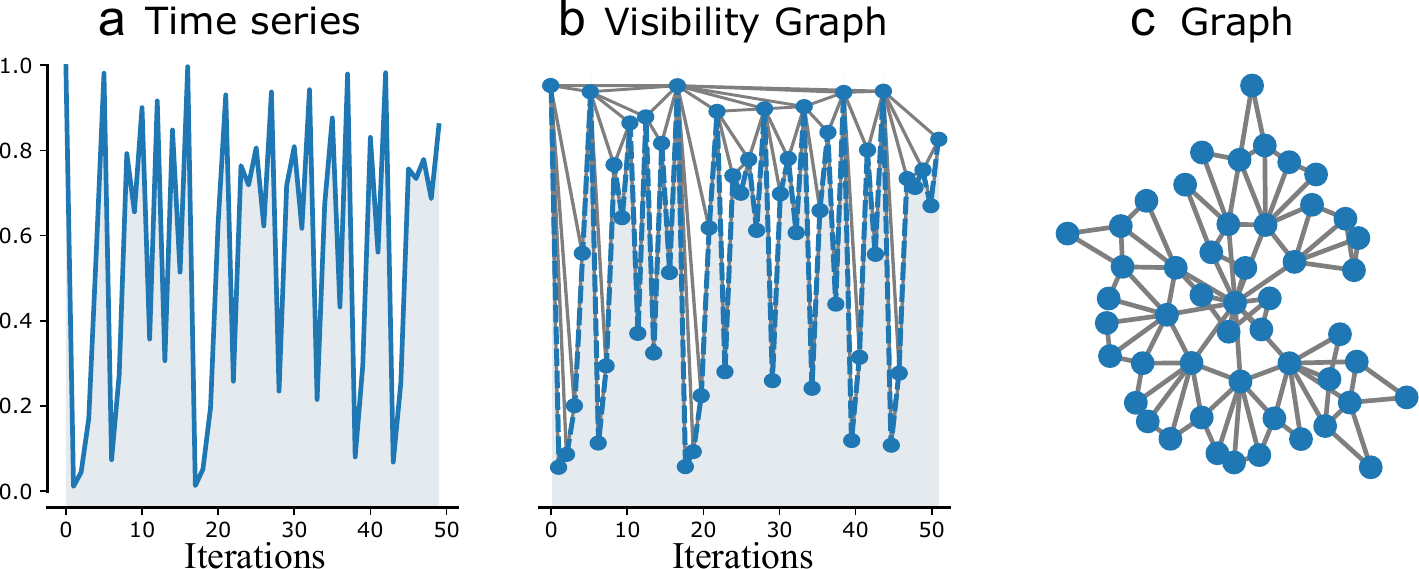}
    \caption{{\bf Network representation of time series.} (a) A Horizontal Visibility Graph (HVG) is a graph representation constructed from time series or one-dimensional data. The HVG is formed by converting time series data into a graph using an algorithm based on the idea of horizontal visibility. (b) Each time series point is treated as a node in the HVG. When a data point at one time step "sees" another data point at a different time step and no other point between them has a higher value, a link is established. (c) The HVG enables the use of graph theory to understand the underlying structure or features of time series data in a variety of domains such as signal processing, economics, and natural sciences. }
    \label{fig:hvg}
\end{figure}

A first step in analyzing the structure of complex systems is often to make a picture of it. The human eye is extremely powerful at picking out visual patterns and allow us to put visualization to work on our studies. A network $G=(V,E)$ is a graphical representation that consist of just two fundamental elements: nodes (or vertices) $v_i \in V$ and pairs of nodes $(v_i, v_j)\in E$ so-called links or edges. Despite this apparent simplicity, networks can express complex patterns by connecting nodes and links in numerous ways. This approach may be beneficial even in systems that do not inherently suggest a network \citep{lacasa2008time} (see Figure \ref{fig:hvg}).\\

Network visualization is, however, only really useful for (a) small networks (up to a few hundred or thousands of vertices), and for (b) sparse networks, that is, when the density of connections is low as not to obscure the main structural features. If there are too many vertices or links, then renderings will be complicated and difficult to understand (such network diagrams are popularly known as 'hairballs'). Many of the networks that interest scientists today have hundreds of thousands or even millions of nodes, which means that visualization is not of much help in their analysis and we need to employ other techniques to determine their structural properties. Here, network theory has developed a large toolkit of measures and metrics that can be useful to understand what data are telling us, even in cases where visualization is not possible. \\

\subsection{Degree}

A convenient notation for mathematical purposes is the adjacency matrix. The adjacency matrix $A$ of a simple network $G=(V,E)$ is the matrix with elements $A_{ij} = 1$ iff $(i,j) \in E$ and $A_{ij}=0$ otherwise. The structure of the adjacency matrix is informative. For an adjacency matrix with no self-edges, the diagonal elements are all zero, while a symmetric matrix shows that the network edges are all bi-directional (so-called 'undirected' edges). In some cases, we would like to represent edges as having a strength, or a weight, or a value attached to them. For example, Internet edges might have weights representing the amount of data flowing along them. In a social network, edges might have weights representing the frequency of interactions. In neural networks, learning processes reinforce specific connections between neurons, and weights might represent the likelihood of signal propagation from one neuron to the next. These weighted networks can be represented by giving the elements of the adjacency matrix values equal to the weights of the corresponding connections, i.e., the volume of traffic carried by a communication link, the number of social interactions, or the strength of a neuronal connection.\\

The most basic structural property of any node $i$ is its degree  $k(i)$ or the number of links connected to it. The degree can be written in terms of the adjacency matrix as follows: 

\begin{equation}
    k_i = \sum_{j=1}^N A_{ij}
\end{equation}
where  $N=|V|$ is the number of nodes. Every edge of an undirected network has two endpoints and if there are $m$ edges in total there are $2m$ ends of edges:
\begin{equation}
    2m = \sum_{i=1}^N k(i)
\end{equation}
or
\begin{equation}
    m = \frac{1}{2}\sum_{i=1}^N k(i) = \frac{1}{2} \sum_{i=1}^N \sum_{j=1}^N A_{ij}
\end{equation}
The above result is often used in the study of networks. Notice that the maximum possible number of edges in an undirected network is
\begin{equation}
{N \choose 2} = \frac{1}{2}N(N-1)
\end{equation}
By dividing the number of edges $m$ by this maximum possible, we obtain the connectance ($\rho$) or density of a network, i.e., the fraction of edges that are actually present~\citep{poisot2014ecological}: 
\begin{equation}
\rho = {{m}\over{N \choose 2}} = {{2m}\over{N(N-1)}} = {{\left<  k \right >}\over{N-1}}
\end{equation}
where $\left<  k \right >$ is the average degree of a vertex in an undirected network. A dense network is one in which the density tends to a constant as $N$ approaches infinity. In this situation, as the network grows in size, the proportion of non-zero entries in the adjacency matrix remains constant. The density of a sparse network, on the other hand, approaches zero as the network size increases, and so the average degree tends to remain constant. Almost all real-world systems are thought to be sparse networks, which has profound implications for our understanding of complex systems \citep{sole2004information}.

\subsection{Percolation}

A system is a set of nodes that are linked together through a chain of intermediate connectors. Nodes that have no connection to the rest of the connected group of species, genes, or people are almost certainly to be removed from the global structure. This raises the question of how the system components were first joined together.\\

\begin{figure}[ht]
    \centering
\includegraphics[width=0.85\textwidth]{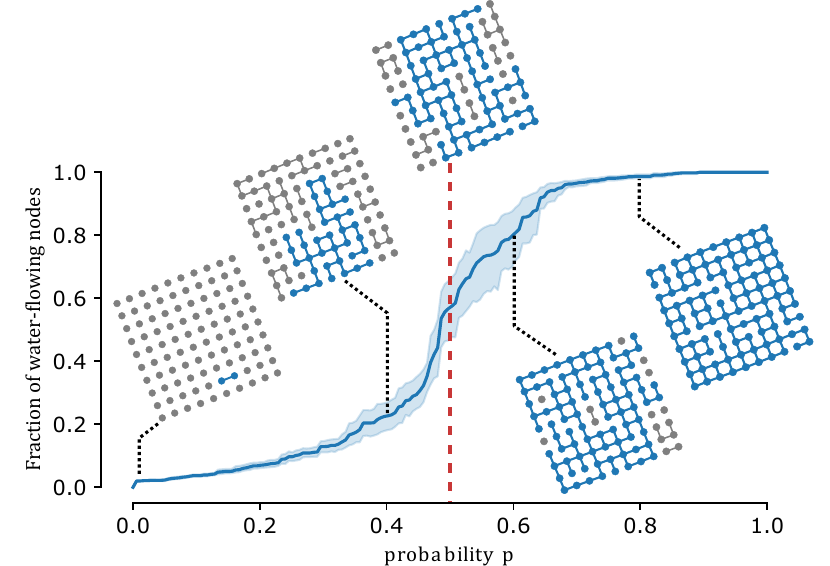}
\caption{{\bf Using a square regular lattice for water transportation.}  In a pipe grid, the probability $p$ (horizontal axis) of opening a pipe (forming an edge) and the total fraction of nodes transporting water (vertical axis) are not linearly related. Nodes on opposing sides of the lattice are part of global transportation routes beyond a critical probability of $p > p_c = 1/2$, even when pipes are randomly opened. Blue and grey circles represent water-flowing and disconnected nodes, respectively.  }
    \label{fig:percolation}
\end{figure}

A simple model provides a broad view of this problem (see Figure \ref{fig:percolation}). Consider a square grid of $L \times L$ water pipes. This spatial network has three types of pipes: (1) pipes that are only open, (2) pipes that are closed, and (3) pipes that are open and may be linked to the left-hand side via a global route. We are interested in the circumstances that allow global routes to appear. Assume that a random pipe on this lattice is opened with probability $p$, with the other pipes remaining independent. Open pipes look separated from one another with low $p$ values. Although the density of open pipes rises linearly with $p$ (on average, we will have $pL^2$ open pipes), the global connectedness does not. The fraction of connected pipes is a non-linear function of the probability $p$. For low $p$, this fraction stays constant until the critical value of $p=1/2$ (the so-called 'percolation threshold') is achieved, at which time the water percolates the grid and a global route connects its opposing sides~\citep{kesten1980critical}.\\

Percolation transitions like these occur in all types of networks. The simplest model of a random network, the Erd{\H{o}}s-R{\'e}nyi graph~\citep{erdHos1960evolution}, provides a null hypothesis for percolation phenomena in disordered systems. The number of nodes $N$ in the random graph is fixed, and each edge is present with probability $p$ and missing with probability $1-p$.  Consider the two extreme scenarios of this model. When $p=0$, each vertex is an island: the network comprises $N$ isolated components, each of which has precisely one vertex. When $p=1$, the random network shows all possible edges, resulting in a single linked component. A 'clique' is the formal term for this configuration in which a network of $n$ nodes is linked to every other node.\\

What happens when probability of connection continuously increases from $p=0$ to $p=1$ in a random network? We may hypothesize that the size of the largest connected component $S(p)$ grows linearly with $p$. But something much more interesting happens. The fraction $S$ of nodes in the largest connected component changes abruptly as the system progresses from disconnected to fully connected. At some critical value of the average degree $z = p(N-1)$, the random network exhibits a phase transition~\citep{sole2011phase}. A large connected component emerges at the critical point $z = z_c = 1$. When $z > z_c$, most nodes are linked in a single graph; however, when $z < z_c$, the system breaks down into many little subgraphs. That is, the random network predicts the emergence of a percolation threshold separating a disconnected phase from another configuration in which most (if not all) components are connected by at least one global route. This prediction applies to all types of networks, as the local structure at the percolation threshold is expected to be a tree with no loops~\citep{callaway2000exact}. However, in real systems, there is also a tendency to short loops~\citep{radicchi2016beyond}, which are also very sparse structures.

\subsection{Clustering}

Random graphs explain the basic features of real-world networks, like the emergence of a large connected component associated with the percolation transition. The random network must be extended to account for other empirical features, particularly when examining local network structures. For example, in social networks~\citep{otte2002social}, the notion of transitivity significantly influences the associations between nodes~\citep{handcock2007model}.The ideal concept of transitivity suggests that connections follow a 'friend of a friend' pattern, often observed in well-connected, fully-linked networks (or cliques). Yet, most real-world networks, being inherently sparse, operate under partial transitivity. For instance, the fact that 'A' is acquainted with 'B' and 'B' with 'C' does not guarantee that 'A' directly knows 'C' in social networks.\\

The clustering coefficient may be used to measure the degree of partial transitivity in real systems. The clustering coefficient $c_i$ of a vertex $i$ is the ratio between the number of edges $e_i$ among its nearest neighbors and its maximum possible value~\citep{newman2018networks}: 

\begin{equation}
c_i= {{e_i}\over{{k_i \choose 2}}} = {{2e_i}\over{k_i(k_i - 1)}}
\end{equation}

This coefficient ranges from 0 to 1.  $c_i = 1$ denotes complete transitivity, i.e., a network composed entirely of cliques. If $c_i=0$, there are no closed triads. This can happen in a number of network types, including trees (which do not have any loops) and square lattices (which have closed loops with an even number of vertices but no closed triads). There is no transitivity or network clustering in random networks because the chance that any two vertices are neighbors is the same, no matter what the nodes are. As a result, in a random network, the average clustering coefficient is $\left < c \right > = p = \left < k \right > / (N-1)$, which goes to zero in the limit of high $N$. Empirical clustering values often range from 0.01 to 0.5, indicating considerable differences between random and real networks~\citep{watts1998collective}.\\

Many real-world networks display an approximate degree dependence, with higher degree vertices having a lower clustering coefficient on average~\citep{maslov2002specificity}. Clustering may also be used to detect so-called "structural holes"\citep{burt2004structural} in the network, or missing linkages between unrelated nodes linked indirectly via a third vertex. Structured holes are negative for transmission efficiency because they restrict the number of alternate pathways through which information may travel. Structured holes, on the other hand, might be advantageous for a central node whose friends lack connections, granting control over information flow between those friends.

\subsection{Motifs}

\begin{figure}[ht!]
    \centering
    \includegraphics[width=0.95\textwidth]{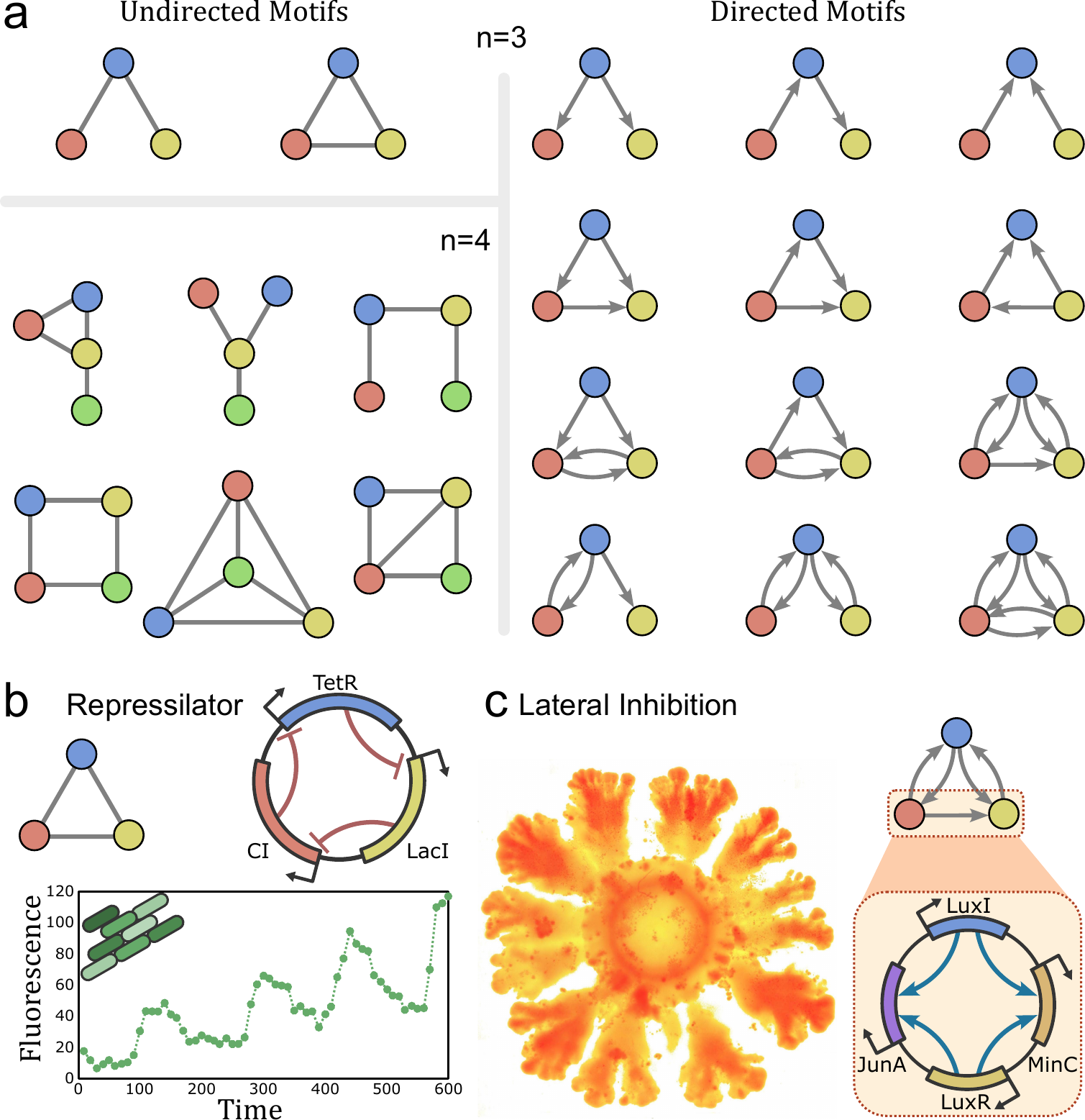}
       \caption{ {\bf Motif analysis in biology}. (a) List of 3 and 4 node undirected network motifs. Directed motifs (right) display a wider range of patterns than undirected motifs (left), resulting in more complex networks of connections.  Synthetic biologists have taken inspiration from these smaller subnetworks to design complex biological circuits.        
       These motifs have proven useful in the implementation of temporal patterns, such as (b) oscillators, e.g., the repressilator circuit~\citep{elowitz2000synthetic}, which features a cascading triad of inhibiting transcriptional regulators (typically TetR, CI and LacI), yielding an oscillation every 150 minutes. The repressilator can be exogenously controlled and owes its robust behaviour despite its simplicity to the underlying transcriptional motif. Robust spatial pattern formation induced by lateral inhibition (c) also hinges on specific design motifs~\citep{duran2021synthetic}, in the example shown a communication-coordinated feedback loop (LuxR-LuxI) promotes cellular adhesion (JunA) while also arresting bacterial growth (MinC), creating regular patterns of cellular density with a characteristic wavelength of 0.2 cm in synthetic {\em E. coli} colonies.}
    \label{fig:motifs}
\end{figure}

Clustering coefficients are instrumental when looking at the local structure of social networks. When applied to directed networks, where imbalanced information, energy, or matter flow predominates, this coefficient is less effective. An alternative lens to explore network architecture involves network motifs~\citep{milo2002network}. These motifs denote compact subgraphs comprising $M$ nodes, interconnected by a subset of links within a larger network with $N > M$ nodes. For each value of $M$, a finite set of variants exists (see Figure \ref{fig:motifs}a). Counting the occurrences of subgraphs within a network generates a frequency spectrum that provides key insights into their role as basic building blocks. Some of these subgraphs occur at considerably greater frequencies than expected in random graphs. To assess the significance of these occurrences, the Z-score provides a statistical metric that quantifies the prevalence of a particular subgraph $\Omega_i$ across  the network~\citep{milo2002network}:

\begin{equation}
z(\Omega_i)= {{ N_{real}(\Omega_i) - \left <  N_{rand}(\Omega_i) \right >} \over {  \sigma (N_{rand}(\sigma_i) )  }}
\end{equation}

where $N_{real}$ is the number of appearances of the subgraph $\Omega_i$ in the network, $\left < N_{rand} \right >$ is the average number of appearances of the subgraph $\Omega_i$ in a large number of randomised networks with the same number of nodes and degree distribution as the original network, and $\sigma$ denotes the standard deviation. A subgraph's Z-score may be positive or negative; $z(\Omega_i) > 0$ indicates that it is greatly overrepresented (it is a motif) in the original network relative to randomized, and $z(\Omega_i) < 0$ indicates that it is severely underrepresented (anti-motif). There are some subgraphs that are overrepresented in the structural and functional networks of the human brain~\citep{sporns2004motifs, mirzasoleiman2011failure}.\\

Although motif analysis is a valuable tool, e.g., when assessing the vulnerability of complex networks~\citep{dey2019network}, its biological relevance in general remains unclear~\citep{sole2006network, knabe2008motifs}. However, synthetic biologists have made extensive use of the motif framework to understand and exploit design spaces in their biological designs. For instance, an exhaustive analysis of feedforward motifs in regulation has revealed that the most common 'band-pass filter' design in evolution is also the most robust one, both in terms of dynamical properties as well as resilience to mutations in their components~\citep{schaerli2014unified}. Similarly, understanding biological motifs underpins our capacity to engineer self-organized criticality \citep{vidiella2021engineering}, associative learning \citep{macia2017synthetic} and more generally, both  temporal~\citep{elowitz2000synthetic} and spatial patterns~\citep{duran2021synthetic} (see Figure \ref{fig:motifs}b,c).\\

Still, it would be unreasonable to directly infer functionality from the presence or absence of specific motifs, given that a specific subgraph might fulfill a variety of roles in different systems. For example, flexible components that modify their mode of operation to replace damaged or absent components might achieve resilient functionality~\citep{tononi1999measures}. According to synthetic models, brain networks are formed to maximize the number and diversity of motifs, as  local structural variability enables an extensive spectrum of functional states~\citep{sporns2004motifs}. Computational research suggests that the degree of synchronization exhibited by a motif is influenced by its structure. Motifs with greater density of links are more likely to synchronize compared with those with fewer connections. 

\section{Evolution}

While random graphs offer insights into network structures, they possess a fundamental limitation in their static nature. Random theory depict networks where all nodes are simultaneously created and remain unaltered, overlooking the dynamic expansion processes prevalent in real-world systems. Take cities, for example. They exhibit a dynamic nature akin to living organisms \citep{weinstock2010architecture, weinstock2011metabolism}. The complex development of urban networks has been studied from the perspective of fractals~\citep{batty1994fractal} and chaos theory~\citep{cartwright1991planning}, since both include self-similar structures as well as sensitivity to initial conditions. This complex city expansion often reflects a combination of distinct developmental patterns: an urban network core surrounded by a grid-like expansion that conforms to the principles of urban planning, while the periphery experiences a more organic and unpredictable growth \citep{valverde2013networks}.  
A static depiction of networks overlooks the complex interplay among historical events, regulatory influences, and selection factors. Embracing dynamical models is crucial to understanding the evolving dynamics of cities~\citep{barthelemy2018morphogenesis} and other complex networks.\\

An open question is to what extent we can define a universal theoretical framework for evolving networks. While basic growth mechanisms, like preferential attachment, contribute to explaining common properties such as the power-law degree distribution, they fall short in capturing the distinctive nature of individual networks. Understanding the evolution of real-world networks necessitates a combination of structural traits beyond local features. 


\subsection{Growth}

Many interesting examples of growing networks are related to evolutionary processes, whether in natural or artificial systems. For example, although technical breakthroughs are frequently the result of purposeful efforts, George Basalla observed that a broad range of inventions cannot be explained solely by intentional processes, implying deep analogies between biology and technology that have been the focus of extensive research~\citep{basalla1988evolution, obrien2000applying, steadman2008evolution}. Furthermore, networks may be used to explore the parallels (as well as the differences) across various evolutionary systems. Historical events both have a significant impact on the course of technology and biology, often resulting in unexpected detours on the path to innovation~\citep{arthur1994increasing}. Biology and technology also depend on differential replication, in which accidental, seemingly random changes are inherited by descendants. Similar to selection forces in evolution, artifacts are also susceptible to external success measures~\citep{mesoudi2021cultural}, such as reliability, performance, novelty~\citep{duran2022dilution, newberry2022measuring}, and even aesthetic preferences.\\

Networks are a promising approach to study the evolution of technology, which lies at the intersection between statistical physics, computer science, and evolutionary biology~\citep{sole2013evolutionary}. The network of patent citations, for example, is a scale-free network, a property also observed in other physical, natural, and man-made systems in which the degree distribution $P(k)$ is a power-law rather than a normal distribution:

\begin{equation}
P(k) = U k^ {-\gamma}    
\end{equation}
where $U$ is a constant, and $2 \le \gamma \le 3$ is the exponent of the power-law \citep{barabasi2016network}. Power law distributions are different from normal distributions in that they do not have a peak at the average degree $\left < k \right >$, and they are more likely to contain extreme degree values.\\

Assuming that the amount of citations is a reliable indicator of a technology's importance, the power-law distribution of citations may be interpreted as an indication of inequality between innovations. But what factors contribute to the popularity of a node? De Solla Price suggested the `cumulative benefit' mechanism for citation networks in 1965, stating that `there is a likelihood that the more an article is cited, the more probable it is to be mentioned later'~\citep{dj1965networks}. This process is similar to the `rich get richer phenomenon' proposed by Yule~\citep{yule1925ii} and Simon~\citep{simon1955class}, as well as the newer concept of `preferential attachment'~\citep{barabasi1999emergence}.\\ 

We can incorporate the mechanism of `cumulative advantage' into a model of growing networks, where the rate of incoming links is proportional to the degree of any node:

\begin{equation}
\Pi (k) \sim k ^ {\beta}    
\end{equation}

where the attachment exponent $\beta$ regulates the strength of link reinforcement. It is possible to demonstrate that the above equation produces a degree distribution that follows a power-law~\citep{yule1925ii, barabasi1999emergence}.  Although this model is often simplistic for reproducing real-world degree distributions, it can be easily expanded. For example, this equation does not take into account existing temporal relationships between inventions. To understand the evolution of technology, we must consider that individual traits are not the unique criteria of success; the environment must also be adequate, since their role may change over time in evolution (e.g., an exaptation~\citep{gould1982exaptation}). Taking into account this, we can define a more realistic model for the growth of patent citations that combines preferential attachment with temporal dependence~\citep{valverde2007topology}:
\begin{equation}
\Pi(k, \tau) \sim k^{\beta}{\tau}^{\alpha-1} \exp (\tau/ \tau_0)^\alpha 
\end{equation}
where $\tau$ is the node age, the exponent $\alpha > 0$ weights how quickly the ageing influences the chance of attachment, and $\tau_0$ is a scale parameter that determines the rightward extension of the ageing curve. We can verify these assumptions in a database of over 3 million patents~\citep{valverde2007topology}, which suggests a complex interplay between structure and environment in the evolution of technology. However, a  power-law degree distribution is a general feature that many systems have, hence it cannot identify what distinguishes a network from others.

\subsection{Modularity}


A modular system is formed by quasi-independent parts that appear integrated within themselves, but also exhibit a certain degree of interdependency among them. Modularity is considered a prerequisite for the adaptation of complex organisms and their evolvability~\citep{schlosser2004modularity}. This is particularly evident in cellular networks, where modularity may be identified at the topological level~\citep{rives2003modular, sole2008spontaneous}. In many situations, modular architectures seem to be linked to functional traits: for example, a group of closely similar proteins may all be involved in cell division or communication~\citep{jonas2011modularity}. While empirical research, e.g.~\citep{erdi2013prediction}, suggests a potential relationship between modularity and functional traits, no universal laws or consistent patterns have emerged in relation to modular structure. A main obstacle is the need to get reliable functional data, which involves the detailed and costly inspection of a large number of nodes.\\

\begin{figure}[hb!]
    \centering
    \includegraphics[width=0.85\textwidth]{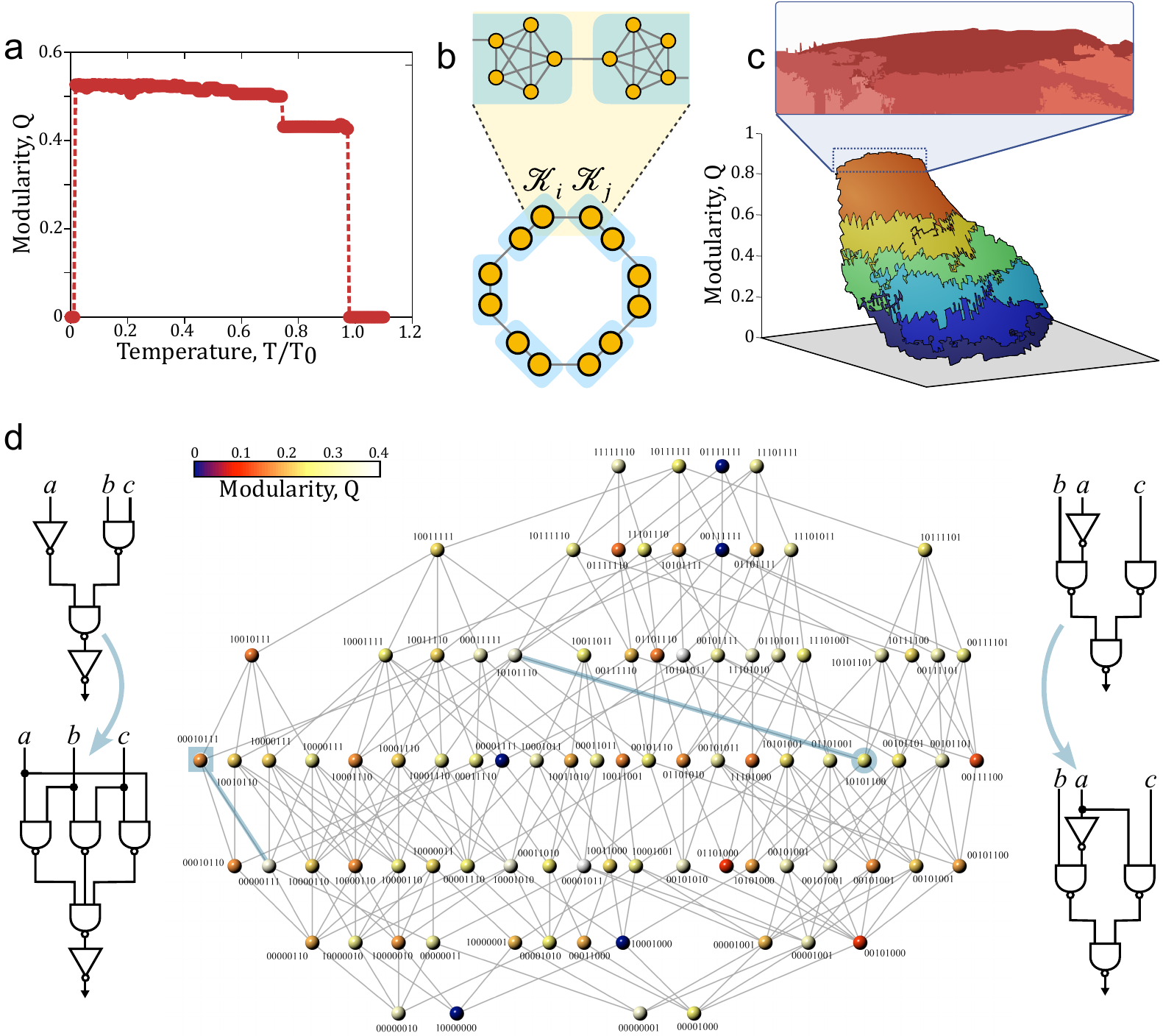}
       \caption{{\bf Limits to modularity.} (a) A simulated annealing search across a broad temperature range finds abrupt modularity transitions separating homogeneous phases. The presence of stable phases suggests how a partial (heuristic) search is likely to provide sub-optimal modularity values. (b) Pairs $(\mathcal{K}_i, \mathcal{K}_j)$ of 5-node cliques are connected by a single link (top) to create a 24-clique ring network (bottom). The best modularity partition, which combines two neighboring cliques, has slightly larger modularity $Q_2 = 0.8712$ than the intuitive partition, which arranges individual cliques on their own, with modularity $Q_1 = 0.8674$. (c) The modularity landscape for this ring network displays a high-modularity plateau of degenerated solutions (inset). (d) A phenotype network (center) for each possible three-input Boolean function, with each node representing one logic function. The functions that are most similar are linked together. The color of a node represents the modularity of the most cost-effective (wire-wise) feed-forward Boolean circuit for implementing that function. A light blue square (left) and a circle (right) highlight two important functions: "majority" and "multiplexor." Evolving a circuit while simultaneously minimizing the number of wires and implementing these target functions inevitably reduces the modularity of its predecessors.  The "breakdown of modularity" refers to the transition from modular ancestors to less modular target circuits (both sides of the panel), as depicted in the phenotypic network (light blue arrows). 
      }
    \label{fig:modularity}
\end{figure}

An active area of research combines statistical physics with computational approaches to develop community detection algorithms that solve this problem~\citep{newman2018networks}. The approach takes into account a decomposition of the graph $G=(V,E)$ into a collection of $s$ subgraphs $C = \{C_i \subset V : 1 \le i \le s\}$ that defines a partition. Because there are so many possible partitions, it is critical to have a reliable index for assessing the degree of connectivity between alike nodes in the network. 
The optimal partition is determined by maximizing the modularity index $Q= Q_{\max} = \max(Q)$ \citep{newman2004finding}. The modularity score is:
\begin{equation}
Q = \sum_{i=1}^s {{{e_i}\over{m}}- {\left ({{d_i}\over{2m}} \right )}^2}
\end{equation}
where $e_i$ is the number of edges in module $i$, $d_i$ is the total degree of nodes in module $i$ and $m$ is the number of edges in the full network. This method is widely used, but it comes with several  limitations (see Figure \ref{fig:modularity}). The so-called 'resolution limit'  \citep{fortunato2007resolution} is a significant constraint that forces the algorithm to integrate small modules whenever the merging gives a positive gain. Let's define the gain $\Delta Q$ as follows:  

\begin{equation}
\Delta Q_{ij} = {{e_{ij}}\over{m}} - 2 \left ( {{d_i}\over{2m}} \right ) \left ( {{d_j}\over{2m}} \right ) 
\end{equation}

where $e_{ij}$ is the number of edges between modules $i$ and $j$. A positive gain $\Delta_Q > 0$ occurs when the number of edges between two subsets of nodes is greater than the expected number in a random graph \citep{good2010performance}:

\begin{equation}
e_{ij} > {{d_i d_j }\over{2m}}
\end{equation}

In sparse or large networks, this value can fall below one ( $e_{ij} < 1$), prompting the merging of weakly linked modules, contrary to our intuition that they should remain separate~\citep{good2010performance}.  The resolution limit results from the modularity equation's null model, which assumes equal probabilities of node connections. This assumption may not hold valid in large or geographically embedded networks, as each node may only have a finite "horizon" within which it may communicate with other nodes \citep{valverde2004internet}. For example, each connection in an intercommunication network has a cost that lowers the likelihood of connecting with another one separated by a large distance \citep{clune2013evolutionary}.\\

Modularity maximization presents an inherent challenge: maximizing the first term of the modularity equation often requires including numerous edges within modules, while reducing the second term necessitates dividing the graph into smaller, low-degree modules.  Finding the optimal balance is a hard computational problem. And as a result, all practical modularity algorithms are based on heuristics, which yield suboptimal partitions~\citep{brandes2006maximizing}. Furthermore, it is unclear how to tune the parameters of the maximization algorithm to obtain the optimal solution. One method involves sweeping the parameter space by varying the temperature in a simulated annealing search \citep{schulke2015multiple}. The network gets clustered into an increasing number of groups and modularity values when the temperature is lowered. The shift from low to high modularity is not smooth, but consists of multiple phases separated by abrupt discontinuities (see Figure \ref{fig:modularity}a). The presence of multiple potential solutions, each with slightly different modularity scores for a given network (Figure \ref{fig:modularity}b), reveals a landscape of local maxima \citep{good2010performance}  (see Figure \ref{fig:modularity}c). To address this problem of degeneration, consensus-based solutions that identify resilient communities are necessary.\\

Evolution cannot invariably produce maximally modular networks when functional and cost constraints are present, which is a more significant limitation than the constraints associated with a mathematical definition of modularity. The so-called "breakdown of modularity" involves a transition from modular to well-connected networks \citep{valverde2017breakdown, godwin2015breakdown}, and it could be related with neurological disorders including Alzheimer's disease or schizophrenia \citep{david1994dysmodularity, alexander2010disrupted}. An extensive search of the space of 3-input logical functions reveals that not all minimal feed-forward Boolean networks (FFBNs) are modular \citep{valverde2017breakdown} (see Figure \ref{fig:modularity}d). Looking at this phenotypic network, we can observe that evolution of circuits such as the multiplexer and the majority function, which are important in both electrical circuits and synthetic biology, is linked to a modularity reduction from an ancestral Boolean function. In these circumstances, the only way to retain modularity is to raise the cost of the circuit, for example, by increasing structural redundancy.

\subsection{Tinkering}

Complex systems emerge from the nonlinear interactions among numerous components, leading to unpredictable behaviors. Modular structures are a frequent component of many complex systems, both in nature and in engineering. In technology, modularity provides significant benefits as it establishes distinct task divisions, aiming to reduce costs and enhance overall reliability~\citep{baldwin2000design}. Herbert Simon postulated that a modular system can adapt more rapidly than one that is not modular \citep{simon1962architecture}.  But whether modularity is a reflection of general evolutionary principles or is exclusive to certain functional features is still an open question. The prevalence of modular designs in both natural and human-made contexts suggests the existence of universal processes, while current heuristic methods for identifying modularity paint a different picture. \\

According to \citep{kashtan2005spontaneous}, networks evolved under "modularly varying goals" (MVG) should also be modular.  A network evolves under MVG when it adapts to changing objectives over time, with each goal having the same subproblems as the preceding one. An example is a logic circuit with multiple parts (modules), each implementing a sub-function needed to perform the overall objective. When objectives change, mutations that rewire these modules rapidly become permanent in the population to meet the new objective. Computational investigations have shown that MVG may greatly speed evolution (resulting in a modular network) as compared to evolution with fixed or fluctuating objectives (leading to a non-modular network). However, it is unclear how many biological settings vary in this modular fashion and if they change often enough to produce modular systems.\\

Modularity has also been suggested to evolve, not because it promotes evolvability in fluctuating environments, but as a by-product of selection that limits the number of links~\citep{clune2013evolutionary}. A more intriguing possibility is that mutation mechanisms, like gene duplication, could create a bias towards modular structures~\citep{sole2008spontaneous}. In the field of technology, makers employ previously accessible innovations as an integral aspect of their work, such as by replicating designs and reusing components acquired from others. Technology, like evolution, evolves mostly by tinkering~\citep{jacob1977evolution}, that is, by combining prior components in novel ways~\citep{arthur2009nature}.  
Even in the absence of selection, a basic network model based on duplication and rewiring can predict not only the empirical degree distribution but also the frequency of motifs in the real systems~\citep{valverde2005network}.\\


\begin{figure}[ht!]
    \centering
\includegraphics[width=0.85\textwidth]{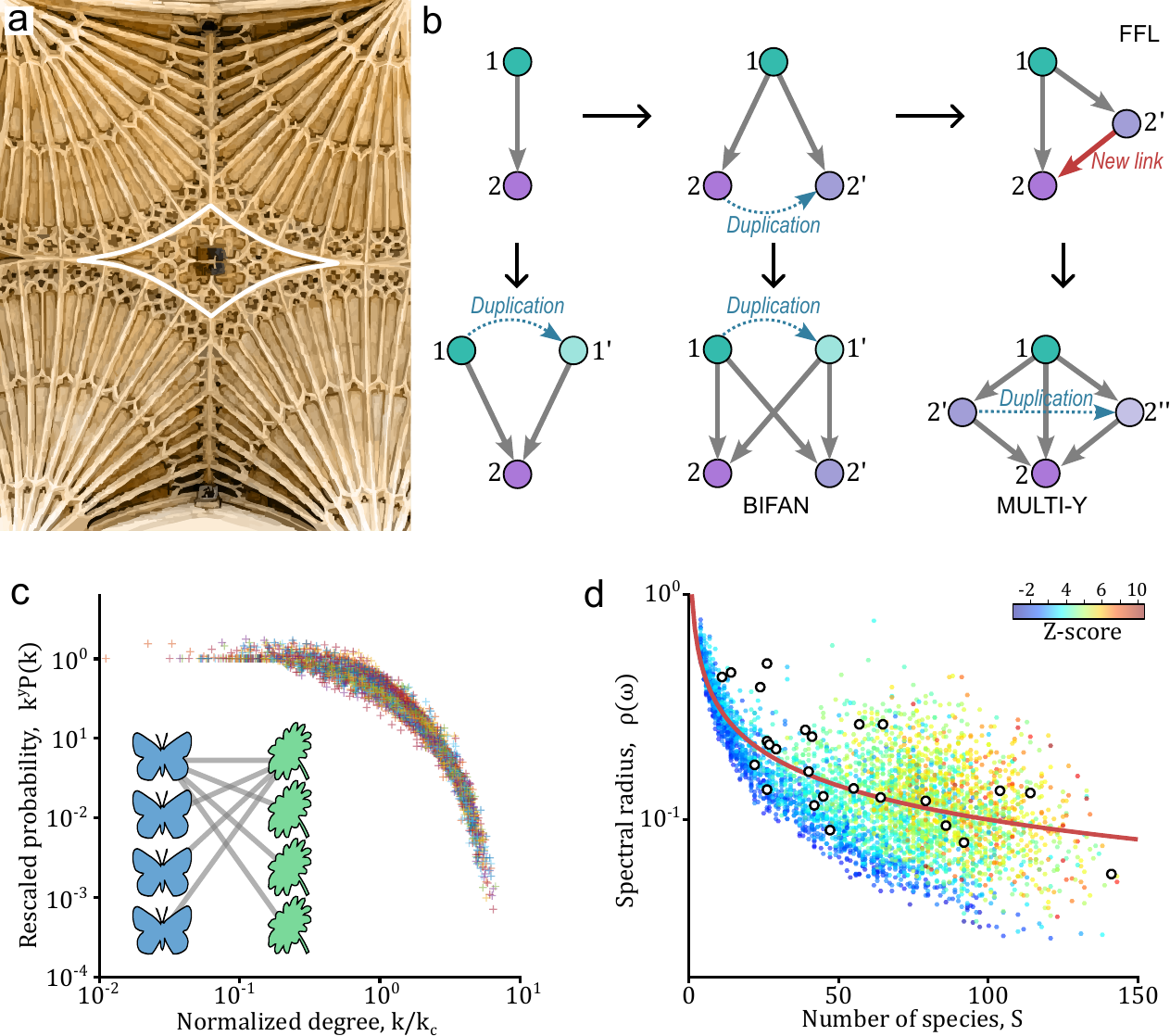} 
       \caption{{\bf Networks as evolutionary spandrels.} Some network properties might be "spandrels," or byproducts or unintended consequences of tinkering rules, rather than the outcome of selection forces. (a) The ceiling of this cathedral shows decorated structures in the middle of four branches (surrounded by a white line). These decorations may be understood as functional essential pieces, although the 
       space (an architectural spandrel) is just an unavoidable result of two arches. (b) We can generate several motifs by copying nodes (blue arrow) or adding new links (red arrow) starting from the simplest motif (top-left). This set of tinkering rules can generate three common motifs (FFL, BIFAN, and MULTI-Y) in directed networks, the (c) truncated power-law behaviour of mutualistic networks (inset) and the (d) relationship between the spectral radius $\rho$ (a robust measure of nestedness, see \citep{staniczenko2013ghost}) and the number of species (S) (color nodes). Open circles correspond to a total of 25 real mutualistic networks \citep{valverde2018architecture}. The red line represents the random network prediction for nestedness, which is based on Wigner's semi-circle rule \citep{arnold1971wigner} and describes the average behavior but not the dispersion of real systems. }
    \label{fig:spandrels}
\end{figure}

Agreement between data and generative models lead us to propose that motif distribution is equivalent to the concept of evolutionary spandrel discussed by Gould and Lewontin~\citep{sole2006network}. A popular biological argument is that structural features (such as modularity) have been selected for the advantages they provide. The logic goes as follows: certain features improve stability, improving the odds of survival; hence, these traits must be evolutionary adaptations. Some features, however, are little more than a `spandrel,' as Gould and Lewontin call them~\citep{gould2020spandrels, koonin2016splendor}. For example, we may observe ornamented structures in the midst of four branches on the ceiling of a cathedral (see Figure \ref{fig:spandrels}a). Although these decorations look like important functional elements, they really reside in triangular spaces known in architecture as spandrels, which are an inevitable outcome of having two arches. Similarly, even in the absence of selection, the overabundance of certain subgraphs in networks may be an unavoidable result of tinkering processes (see Figure \ref{fig:spandrels}b).\\

\subsection{Nestedness}

A widespread and intriguing pattern in ecological networks is nestedness, which is characterized by the tendency of low-degree species to interact with a subset of highly connected species~\citep{bascompte2003nested}. The canonical interpretation of nestedness in ecological networks is linked to the likelihood observing a particular pattern of interactions compared to a null model~\citep{weitz2013phage}. This analysis is traditionally carried out on topological matrices (meaning that they only account for the presence or absence of links). Such matrices are typically nested, but it has been pointed out~\citep{staniczenko2013ghost}, that a better characterization of these webs using link weights reveals that only a small fraction of them exhibit nestedness~\citep{timoteo2022tripartite}, suggesting that a truly meaningful assessment of nested patterns requires a weighted interaction matrix.\\

Edges in a nested network are organized in such a way that specialists interact with a subset of the species whom generalists interact with. This definition has been recently extended to quantitative networks using spectral graph theory~\citep{staniczenko2013ghost,valverde2018architecture} (see Figure \ref{fig:spandrels}d). Formally, nestedness is a systematic arrangement of non-zero entries in the adjacency matrix. This pattern can be clearly identified by computing the so-called overlap and declining fill (NODF\citep{almeida2011straightforward}). For the bipartite graph $G=(P,M,E)$ with biadjacency matrix $B=[B_{ij}]$:

\begin{equation*}
NODF(G) = \frac{1}{K} \left [ \sum_{i,j=1}^{N_P} \left ( \theta (q_i - q_j) \frac{\sum_{k=1}^{N_M} B_{ik} B_{jk} }{q_j} \right ) + \sum_{k,l=1}^{N_M} \left ( \theta (r_k - r_l) \frac{\sum_{i=1}^{N_P} B_{ik} B_{il}}{r_l} \right ) \right ]
\end{equation*}

where $K= [ N_P (N_P -1 ) + N_M (N_M - 1)]/200$ is a normalisation constant to ensure that $0 \le NODF \le 100$, $N_P=|P|$, $N_M=|M|$, $\theta$ is the Heaviside function with $\theta(0)=0$, and $q_i$ and $r_j$ are the degrees of nodes $i \in P$ and $j \in M$, respectively. A high NODF value implies that certain species' interactions are a subset of other generalist species' interactions (and so the network demonstrates nestedness), whereas a low value indicates clustering (thus implying modular structures).\\

Following an adaptationist view of naturally evolved systems, it has been argued that the presence of nested patterns is a consequence of underlying selection processes that reduce competition relative to the benefits of facilitation, increasing biodiversity~\citep{bascompte2007plant} and food web persistence~\citep{levine2017beyond}. Supporting evidence for this view are analytical approaches to generalized Lotka–Volterra equations with different functional responses. However, recent papers have challenged this view~\citep{timoteo2022tripartite}. Instead, it has been argued that nestedness is likely to be a consequence (rather than a causative property) of biodiversity, in particular of the heterogeneous distributions of connections~\citep{dunne2002network,staniczenko2013ghost} (see Figure \ref{fig:spandrels}c). Furthermore, null models~\citep{valverde2018architecture} have demonstrated that nestedness can be easily generated from generative processes; this suggests that tinkering models, which rely on basic copy and rewire processes, generate nested structures, thereby implying that nestedness is an evolutionary spandrel~\citep{gould2020spandrels,valverde2018architecture}. Thus, validation of empirical nested patterns should take into account the predictions of tinkering models as a baseline to assess whether the observed patterns are meaningful, i.e. deviating from null expectations (see Figure \ref{fig:spandrels}d). 

\section{Beyond Networks}

\begin{figure}[!ht]
    \centering
    \includegraphics[width=0.95\textwidth]{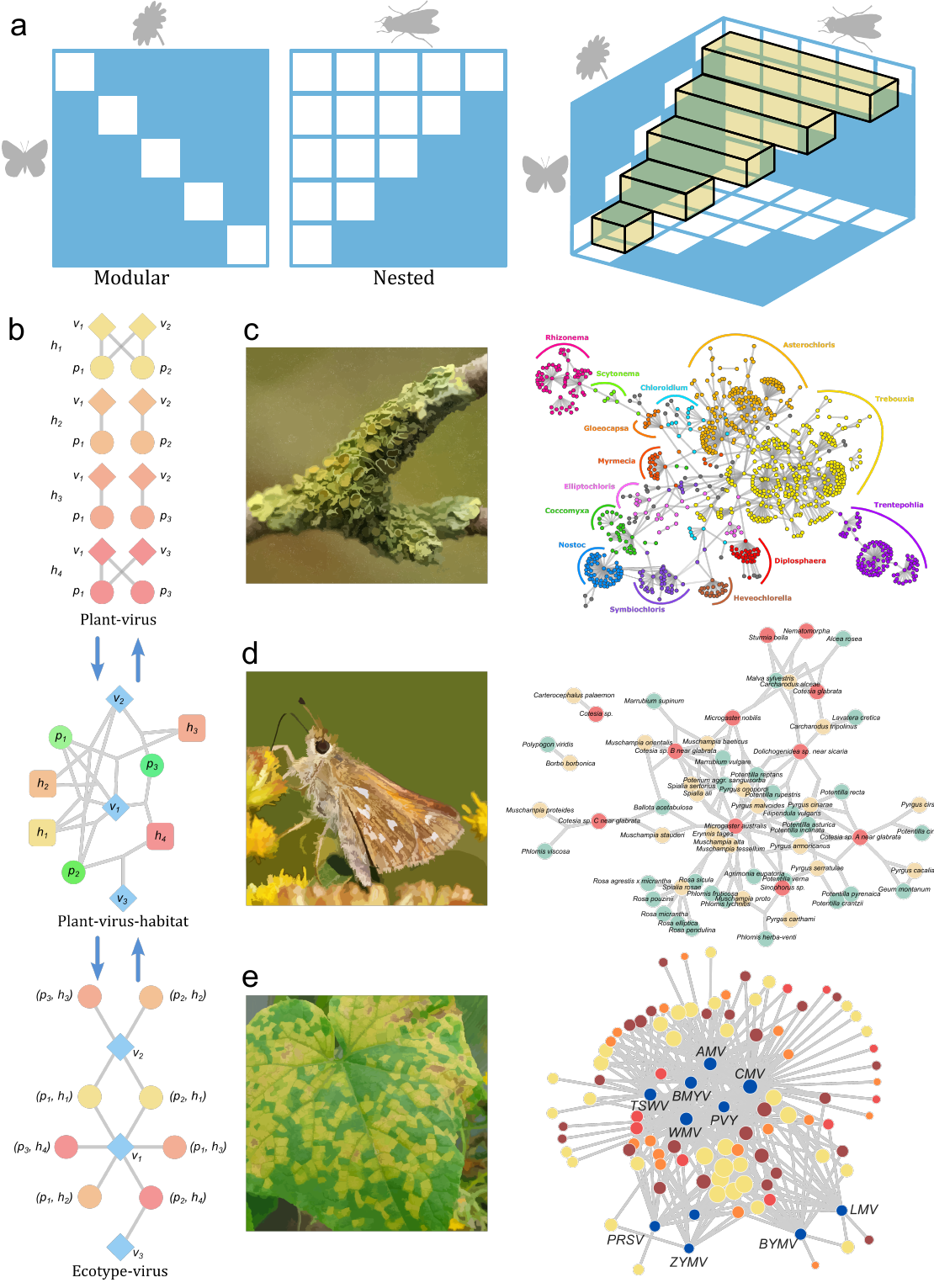}
       \caption{ ...  Hypergraphs can also be reconstructed from `partial' information, such as separate bipartite networks (b). Here, two bipartite networks are integrated into a hypergraph containing plants, viruses and biomes~\citep{valverde2020coexistence}. Many systems are amenable to a hypergraph representation: from lichen symbionts that contain many fungi-algal-bacterial relations~\citep{duran2023composition} (c), plant-butterfly-parasite interactions~\citep{toro2022butterfly} (d) and plant ecotype-parasite infection~\citep{valverde2020coexistence} (e).\\}
    \label{fig:hypergraphs}
\end{figure}

Networks offer an efficient means of capturing both local and global structural features, serving as a valuable framework for investigating the dynamic properties of systems. However, their scope is limited to such pairwise interactions. In contrast, hypergraphs provide a solution to this limitation by allowing edges to connect any number of vertices (see Figure \ref{fig:hypergraphs}b), rendering them exponentially more powerful and expressive in describing real-world interactions. 
Following previous notation, a hypergraph $H=(V,E)$ is a graph consisting of nodes (or vertices) $v_i \in V$ and and hyperlinks (or hyperedges) which can span any number of nodes $(v_i ... v_j)\in E$.\\

For instance, hypergraphs are indispensable when pathogens exhibit varied infection behaviors based on the presence or absence of multiple disease vectors~\citep{rizzoli2019parasites} (see Figure \ref{fig:hypergraphs}d), or in cases of symbiotic relationships involving more than two species~\citep{miransari2011interactions, duran2023composition}. In other ecological systems, conventional depictions of predatory interactions, such as those in food webs, typically involve only two species. However, research has repeatedly demonstrated that many ecological interactions are of higher order~\citep{abrams1983arguments}. This means that they can involve more than two species~\citep{golubski2016ecological}, incorporate external ecological factors that influence pairwise interactions~\citep{levine2017beyond,valverde2020coexistence} (see Figure \ref{fig:hypergraphs}e), or are driven by density-dependent behavior (such as when prey change their behaviour leading to their predators switching to other more suitable preys). In these cases, a network formalism may fall short in capturing the essential structural properties that drive emergent dynamical features. Hypergraphs become a necessary tool for making accurate predictions in systems with higher order interactions, due to their enhanced expressiveness and flexibility, rendering them highly valuable in complex systems research.\\

Given that the structure of ecological networks has been linked to patterns of extinction and robustness~\citep{bornholdt2000robustness}, how does a hypergraph perspective address these views? It has long been proposed that a modular network structure prevent the spread of perturbations (such as secondary extinctions) in ecological networks, with modules acting as firewalls against the propagation of extinction cascades~\citep{stouffer2011compartmentalization, olesen2007modularity}. This suggests that modular patterns should be more resilient and, consequently, preserved in evolutionary processes. Conversely, mutualistic networks tend to exhibit a nested rather than modular pattern, which has also been associated with increased resilience and productivity in various ecological systems~\citep{bascompte2007plant}. This assessment goes beyond mere conjecture; Sanders and colleagues demonstrated through field experiments with plant-insect communities that more complex food webs can indeed buffer against the effects of species loss~\citep{sanders2018trophic}.\\

Traditionally, modularity and nestedness were considered mutually exclusive~\citep{staniczenko2013ghost,fortuna2010nestedness} and imbued ecological networks with different dynamical properties~\citep{bascompte2007plant}. Various network models have emerged, attempting to account for their independent origins. However, recent research in hypergraphs has revealed that when modularity and nestedness inhabit distinct dimensions of high-order interaction graphs, or hypergraphs, they can coexist~\citep{valverde2020coexistence}. In this way, a projection of a tripartite hypergraph into a network (by collapsing the adjacency matrix from 3 to 2 dimensions) can recover either structural pattern (see Figure \ref{fig:hypergraphs}a).\\

\begin{figure}[b!]
\noindent\textcolor{mylinkcolor}{\rule{\textwidth}{.1mm}}
    {\textbf{Fig. 7 }
  {\bf Hypergraphs examples and bipartite projections.} A tripartite hypergraph of host-parasites-habitat can be decomposed or projected into host-parasite or host-plant bipartite networks (a). These projections can display distinct structural patterns such as the host-plant network being modular (left) and the host-parasite being nested (right). ...}
\end{figure}

\clearpage
\vfill
\vfill
\clearpage

Hypergraphs can also help researchers integrate different types of interactions in a single graph object. For instance, to better understand the compounding effect of human laws in biodiversity and biological interactions~\citep{wells2007use,epstein2016legal}; the complex relationships between landscape, species and ecosystem services~\citep{felipe2020land}; how anthropogenic climate change drives the reorganization of ecological networks~\citep{gomez2019climate}. With hypergraphs it also becomes possible to incorporate spatial or temporal dimensions to ecological networks~\citep{valverde2020coexistence}, allowing us to address when structural patterns are driven by seasonal variation, or when interactions are guided by `ecotypes' (see Figure \ref{fig:hypergraphs}e, i.e. when species change their morphological features depending the biome they inhabit~\citep{jackson2017topological}, thus altering the suite of interactions they partake in). However, although many of the structural metrics mentioned in this chapter also exist for hypergraphs~\citep{battiston2020networks}, the field of empirical research in hypergraphs is still pretty much in its infancy, and further examples and mechanistic models will be necessary to bring them up to par to the ubiquity of network analysis.\\

\section{Discussion}\label{sec:discussion}

Biological complexity results from adaptations and shifts to novel features, taking place at various scales. Not all network characteristics, however, are the outcome of direct selection processes; rather, they can unexpectedly emerge from internal dynamics, akin to evolutionary spandrels. Scale-free degree distributions, motif frequencies, and even nestedness may have emerged as accidental outcomes of non-adaptive processes, such as tinkering rules. Simple growth models, however, are insufficient to explain how network complexity has evolved over time. Exogenous and endogenous mechanisms both contribute to the emergence of complexity.\\

Not all innovations lead to more complexity; the choice of which ones persist is ultimately based on environmental conditions and cost-benefit analyses. In technology, software engineers try to manage complexity by rationally subdividing a system into modules that interact in a neat and clear way~\citep{baldwin2000design}. External sources of complexity, like performance constraints, can entangle the original design, which soon becomes much less modular~\citep{valverde2017breakdown}. Software developers actively oppose complexity in this situation because unwanted complexity makes the system much more challenging to understand and maintain, often leading to entangled and monolithic systems~\citep{eick2001does}. Another source of complexity is the design principle that maximizes reuse of existing components. The merging of an existing system with an external one, e.g., a third-party library, results in complex co-evolutionary dynamics between different software projects \citep{valverde2007crossover}. \\

\begin{figure}[ht!]
    \centering
    \includegraphics[width=0.6\textwidth]{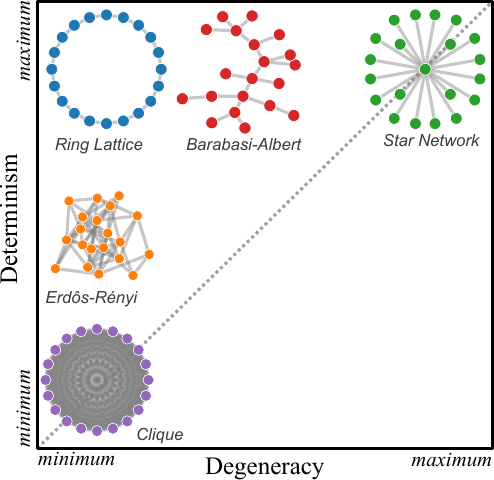}
       \caption{{\bf Network morphospaces.} Degeneracy $\langle H(W_{i}^{out}) \rangle$ and determinism $H( \langle W_{i}^{out} \rangle)$ display a fundamental trade-off which confines networks to a specific domain of this design space~\citep{klein2020emergence}. Here, $H(X)=-\sum_n P(x_i)\log_2 P(x_i)$ represents the Shannon's entropy of a given distribution $X$, and the distribution $W_{i}^{out} = \{w_{ij}\}/{\sum_j w_{ij}}$ is the normalized set of weights from node $i$ to every other node $j$. These weights can be understood as the distribution of probabilities a random walker might take when visiting node $i$. Regular structures (star, clique), lattices (ring) and typical networks produced by standard generative models (Erd{\H{o}}s-R{\'e}nyi and Barab\'asi-Albert) are shown in this space. A forbidden region in the bottom right corner of the morphospace is uninhabited by any real network.}
    \label{fig:morphospace}
\end{figure}

We can use network morphospaces to characterise the vast landscape of network designs and how external forces and internal constraints drive structural diversity. These are representations of design spaces, where different networks are located according to some well defined metrics. Typically, the morphospace dimensions are chosen because they reflect some deeper insight into the evolutionary process or constrains that limit what is possible in these systems, i.e. they often display trade-offs between them. One such example is the network degeneracy-determinism morphospace (see Figure \ref{fig:morphospace}). This space quantifies the information contained in the topology of a network and can be used to signal the presence of higher informative scales. Here, determinism corresponds to the degree of determinism or certainty associated to a random walker traversing the network in every node, and the second dimension captures the system's degeneracy or entropy in weight distributions~\citep{klein2020emergence}. The features of this morphospace have implications for the dynamical properties of brain activity~\citep{hoel2016can}.\\

In this chapter we have explored a wealth of tools, metrics and morphospace representations that allow us to map and navigate complex systems. Understanding complexity not only gives us insights into the generative processes that underlie biological and technological systems, but also informs the software and biological engineers that want to manage the complexity of their designs. Stephen Hawking's prediction of a 'century of complexity' seems to be spot on with our current reality, which is distinguished by extraordinary technological advances contrasted against profound social and planetary challenges. In this age of complexity, our analysis of network architectures barely scratches the surface. Underlying there is a vast, undiscovered terrain, just like the bigger complexities awaiting discovery and comprehension in the world around us.\\

\begin{acknowledgement}

This chapter is based on S. V.'s Introduction to Complex Networks course notes from the University of Val\`encia's Master of Computational Systems Biology program, which is led by Professor Juli Peret\'o (2014-). S.V. \& B.V. thanks Michael Stich for his kind invitation to participate in the 3rd Workshop of Nonlinear Dynamics in Biological Systems. S.V. is supported by the Spanish Ministry of Science and Innovation through the State Research Agency~(AEI), grant PID2020-117822GB-I00 /AEI/10.13039/501100011033. B.V. and S.V. are supported through the 2020-2021 Biodiversa and Water JPI joint call under the BiodivRestore ERA-NET Cofund (GA N°101003777) project MPA4Sustainability with funding organizations: Innovation Fund Denmark (IFD), Agence Nationale de la Recherche (ANR), Fundação para a Ciência e a Tecnologia (FCT), Swedish Environmental Protection Agency (SEPA), and grant PCI2022-132936 funded by MCIN/AEI/10.13039/501100011033 and by the European Union NextGenerationEU/PRTR. S.D-N. is supported by the Beatriu de Pinós postdoctoral programme, from the Office of the General Secretary of Research and Universities and the Ministry of Research and Univertisites (2019 BP 00206) and the support of the Marie Sklodowska-Curie COFUND (BP3 contract no. 801370) of the H2020 programme. We acknowledge the support of the network PIE-202120E047-Conexiones-Life.
\end{acknowledgement}

\end{document}